# Dramatic reduction of surface recombination by *in-situ* surface passivation of silicon nanowires

*Yaping Dan[†], Kwanyong Seo[†], Kuniharu Takei[‡], Jhim H. Meza[†], Ali Javey[‡], and Kenneth B. Crozier[†*]*

[†]School of Engineering and Applied Sciences, Harvard University, Cambridge, MA 02138
[‡]Department of Electrical Engineering and Computer Sciences, University of California, Berkeley, CA 94720

*Corresponding email: kcrozier@seas.harvard.edu

**RECEIVED DATE (to be automatically inserted after your manuscript is accepted if required according to the journal that you are submitting your paper to)**

ABSTRACT Nanowires have unique optical properties[1-4] and are considered as important building blocks for energy harvesting applications such as solar cells[2, 5-8]. However, due to their large surface-to-volume ratios, the recombination of charge carriers through surface states reduces the carrier diffusion lengths in nanowires a few orders of magnitude[9], often resulting in the low efficiency (a few percent or less) of nanowire-based solar cells[7, 8, 10, 11]. Reducing the recombination by surface passivation is crucial for the realization of high performance nanosized optoelectronic devices, but remains largely unexplored[7, 12-14]. Here we show that a thin layer of amorphous silicon (a-Si) coated on a single-crystalline silicon nanowire (sc-SiNW), forming a core-shell structure *in-situ* in the vapor-liquid-solid (VLS) process, reduces the surface recombination nearly two orders of magnitude. Under illumination of modulated light, we measure a greater than 90-fold improvement in the photosensitivity of individual core-shell nanowires, compared to regular nanowires without shell.

Simulations of the optical absorption of the nanowires indicate that the strong absorption of the a-Si shell contributes to this effect, but we conclude that the effect is mainly due to the enhanced carrier lifetime by surface passivation.

Considerable interest exists for nanowire-based solar cells[2, 8, 15-18] because of their potential to achieve high energy conversion efficiency at low cost. However, due to the large surface-to-volume ratio, nanowires have a high surface recombination rate. This shortens the carrier lifetime by 4-5 orders of magnitude[9], resulting in the efficiency of nanowire solar cells usually only a few percent[7, 8, 10, 11]. The surface passivation of planar optoelectronics is a technology that was developed over the course of decades[19, 20]. Nanowire surface passivation, however, is an even more challenging task due to their small size and the fact that multiple facets with different crystalline orientation are exposed. In this Letter, we demonstrate a highly effective means for surface passivation of silicon nanowires, using a thin layer of amorphous silicon formed in-situ during nanowire growth. Excitingly, the experimental results obtained by near-field scanning photocurrent microscopy indicate a ~100-fold reduction in surface recombination. The quenched surface recombination prolongs the carrier lifetime and consequently increases photosensitivity or energy conversion efficiency when the nanowires are used as photodetectors or solar cells. Experimental measurements indicate a ~90-fold increase in the photosensitivity of passivated nanowires, as compared to unpassivated ones. This has major implications for solar energy conversion devices.

Images of transmission electron microscopy (TEM) and energy dispersive X-ray spectroscopy show that the nanowires have a single-crystalline core and a ~10 nm thick amorphous silicon (a-Si) shell (Fig. 1). The core-shell SiNWs are mostly 40 - 60 μm long and 50 - 90 nm in diameter. The amorphous shell is formed *in-situ* during the nanowire synthesis in the Au-catalyzed vapor-liquid-solid (VLS) process that has been widely used for the growth of SiNWs without shells[21-23] (see the supplementary information Fig. S1 for possible growth mechanisms of the amorphous shell). The core-shell nanowires are p-type, confirmed by the result of transimpedance measurement shown in the inset of Fig. 2a. The boron doping concentration in the SiNW is estimated to be ~$10^{18}$ cm$^{-3}$ based on four-probe measurements.

To quantify the effect of surface passivation in our core-shell nanowires, we extract the surface recombination velocity *S* by the following equation (see the supplementary information Fig. S2):

$$\frac{1}{\tau_{eff}} = \frac{1}{\tau_b} + \frac{4S}{\phi} \quad (1)$$

where $\tau_{eff}$ is the effective carrier lifetime, $\phi$ the nanowire diameter, and $\tau_b$ the carrier lifetime in bulk Si of the same impurity concentration. The left side of the equation is the effective recombination rate which is equal to the addition of bulk (first term on right side) and surface recombination rate (last term on right side). While the nanowire diameter is measured directly by atomic force microscopy (AFM), we obtain the effective lifetime $\tau_{eff}$ from the minority diffusion length $L_{diff}$, since the two are related by $L_{diff} = \sqrt{D\tau_{eff}}$ where D is the diffusion coefficient. To find $L_{diff}$, an electrically neutral region in the nanowire is created by making two Schottky contacts to the wire, as shown in Fig. 2b. Under both positive and negative bias, one of the two Schottky barriers is reverse biased (black line in Fig. 2a). The majority of the applied voltage drops over the reverse biased barrier, with only a negligible part dropped over the nanowire. A ~100 nm illumination spot from a near-field scanning optical microscopy (NSOM) tip is used to locally generate excess minority carriers in this quasi-neutral region. These carriers diffuse and recombine in the absence of an external electric field, leading to the decay in their population. Only a portion of the excess minority carriers generated by the illumination spot therefore reach the reverse biased barrier and are collected as photocurrent. From Fig. 2b, it can be seen that photocurrent increases exponentially as the distance between the illumination spot and barrier decreases. We extract the diffusion length $L_{diff}$ from this exponential relationship between the photocurrent and the illumination spot location.

As discussed, the ~100 nm illumination spot is created by focusing a laser beam into a near-field scanning optical microscope (NSOM) tip [24-26]. This tip is raster scanned over the nanowire in contact mode, enabling the topography (i.e. AFM image) and photocurrent map to be simultaneously obtained, as shown in Fig. 2c and d, respectively. The bright spot in Fig. 2d corresponds to the large photocurrents occurring near the reverse biased Schottky barrier (bottom electrode of Fig. 2c). It can be seen from Fig. 2e that the photocurrent decays exponentially with distance from the illumination spot. From the decay rate, the diffusion length of the 77 nm diameter core-shell wire is found to be 410 nm. Variation in the external voltage modulates the depletion region of the Schottky barrier and changes the carrier collection efficiency as well as the absolute value of photocurrent. However, the photocurrent decay slope, and hence the diffusion length, remain constant at different voltages (Fig. 2e), because the external voltage does not affect the diffusive nature of excess carriers in the neutral region. According to *equation (1)*, smaller nanowires have larger surface recombination rates, leading to shorter diffusion lengths, as observed in Fig. 2f.

We plot diffusion lengths against the nanowire diameters in Fig. 3. For comparison, the diffusion lengths of regular nanowires without shell are also measured (green stars, see the supplementary information Fig. S3 for details), which are consistent with the previous results from ref.[9] (red dots). It can be seen that the core-shell nanowires (black squares) have diffusion lengths 7 - 11 times greater than those without shell. Using *equation (1)* and $L_{diff} = \sqrt{D\tau_{eff}}$, we find two curves (the dashed lines) that fit the two groups of data very well with surface recombination velocities $S$ of $4.5\times10^3$ cm/s and $3\times10^5$ cm/s for the core-shell nanowires and the regular ones, respectively. The other two solid lines in Fig. 3 plot the diffusion length vs diameter curves for two other choices of $S$ ($3\times10^3$ cm/s and $1\times10^4$ cm/s). For the plotted curves, we use a constant diffusion coefficient $D$ (= 4.8 cm$^2$/s with mobility $\mu$ ~200 cm$^2$/Vs) for both types of nanowires, since the carrier mobility is mainly limited by the scattering of majority impurities for nanowires with a doping concentration ~$10^{18}$ cm$^{-3}$ as in our case[9, 27] and therefore is largely independent of the surface scattering[9]. The reduction in surface recombination velocity indicates that the a-Si shell has reduced the density of surface states on the crystalline core surface nearly two orders of magnitude.

The quenched surface recombination prolongs the carrier lifetime in nanowires and consequently increases the concentration of photon-excited excess charge carriers ($\Delta n = G \times \tau$ where $G$ is the carrier generation rate and $\tau$ the minority carrier lifetime) when the nanowires are illuminated. The higher concentration of excited carriers will lead to higher photosensitivity or energy conversion efficiency when the nanowires are used as photodetectors or solar cells. For experimental simplicity, we investigate the photosensitivity of Ohmically contacted nanowires by measuring the photocurrent $I_{ph}$ which is given by:

$$I_{ph} = e \times \Delta n \times \mu \times E \times A \quad (2)$$

where $e$ is the electron charge, $\Delta n$ the excited minority carrier concentration that is given by $\Delta n = G \times \tau$ as above, $\mu$ the minority carrier mobility, $E$ the electric field in the nanowire, and $A$ the nanowire cross-sectional area. The only extrinsic parameter in *equation (2)* is the electric field $E$ (= $V_{in}/L$, Fig. 4a), and this can be accurately measured using the four probe measurement technique. The device is uniformly illuminated with light from a monochromator modulated on/off at a frequency of 1.8 kHz by a chopper. The AC component of the photocurrent is then extracted with a lock-in amplifier, as shown in Fig. 4a. However, e*quation (2)* will not necessarily maintain its linearity as a function of both the generation rate $G$ (proportional to illumination intensity) and the electric field $E$, unless $\tau$ and $\mu$ remain intrinsic constants independent of $G$ and $E$. We find that, for the core-shell nanowire device, the photoconductance is linear with the illumination intensity up to ~25 W/m$^2$ (Fig. 4b), while the photocurrent linearly increases with the electric field over the range explored in our experiments (inset of Fig. 4b). Under these circumstances, $\tau$ and $\mu$ remain as intrinsic constants. Figure 4c shows that the AC photosensitivity (photoconductivity per unit illumination intensity) as a function of illumination wavelength ranging from 400 nm to 900 nm for a core-shell nanowire and a nanowire without shell that are 90 nm and 130 nm in diameter, respectively. The absorption cross-sections of these two wires (on a quartz substrate) are calculated by the finite difference time domain (FDTD) method and plotted in Fig. 4c as solid lines. The absorption cross sections $A_c$ have units of length (rather than area). If an infinitely-long nanowire is illuminated by a plane wave of intensity $I$, then the power absorbed in a length $l$ of the nanowire is given by $I \times A_c \times l$. These lines represent the case of unpolarized illumination, and are obtained by averaging the spectra obtained for TE and TM illumination (Fig. 4a). It can be seen that the measured photosensitivity and simulated absorption cross-section follow very similar trends with the two nanowires each having one or more peaks due to leaky mode resonances[28, 29]. These modes can be found by solving the Maxwell's equations at appropriate boundary conditions for classical optical fiber waveguides, as described previously[28]. Each of the leaky modes can be characterized by an azimuthal mode number, m and a radial order number, n. In this way, the modes can be termed as $TM_{mn}$ or $TE_{mn}$. We plot the configuration of electric field intensity of TM mode resonances in Fig. 4d since they dominate the absorption in nanowires. The core-shell nanowire and the regular nanowire each have a $TM_{11}$ mode resonance at ~ 530 nm and ~670nm, respectively, leading to a broad bump in their absorption spectra. The corresponding photoconductance peaks both blueshift, a phenomenon that was also observed by others[28], though the cause is not perfectly understood. The 130nm regular nanowire has a second bump at ~ 482nm which consists of a main peak at 482nm ($TM_{21}$) and a shoulder at 466nm ($TM_{02}$). The existence of the small shoulder is evident in the asymmetry of the bump. The configuration of electric field intensity shown in the third figure of Fig. 4d is a mix of these two resonances since they are close to each other. The third bump of the regular nanowire at ~406nm is due to the resonance of $TM_{12}$ mode.

The largest photosensitivity for both the core-shell nanowire and the nanowire without shell occurs at wavelengths around $\lambda \approx 490$ nm, at which the core-shell nanowire has a photosensitivity over 90 times greater than the regular nanowire. The FDTD simulations predict that the amorphous shell increases the absorption cross-section by ~ 1.3 times (black and red solid lines of Fig. 4c). We therefore conclude that the remaining factor of ~70 ($\approx$ 90/1.3) in photosensitivity improvement comes from the enhanced carrier lifetime due to surface passivation. This is in fair agreement with the data in Fig. 3. Interestingly, the FDTD simulations predict that the larger absorption of the core-shell nanowire originates almost entirely from the 10 nm amorphous shell (Fig. 5a). This is mainly due to two reasons. First, the electric field within the nanowire has two intensity peaks in the a-Si shell, as shown in the $TM_{11}$ mode plot (Fig. 4d). Secondly, a-Si absorbs light in visible range more strongly than sc-Si[30, 31]. The strong absorption in the a-Si shell generates a large number of charge carriers inside the shell. We note that the absorption cross-section peak at ~530 nm of the core-shell nanowire overlaps approximately with its photosensitivity maximum, though there is an offset. This means that the excited carriers in the a-Si shell are also efficiently collected as photocurrent. However, bulk a-Si has a carrier diffusion length as short as ~50 nm[32], while the diffusion length in the core-shell nanowire is measured to be greater than 400 nm (Fig. 3). It is unlikely that the excited carriers transport in the amorphous shell even though it is in electrical contact with microelectrodes (a-Si in the contact region is not removed before metal evaporation). We therefore conclude that most of the carriers generated in the shell diffuse into the core, where they are then transported to the electrodes as photocurrent (Fig. 5b). This transport is efficient due to high mobility of the sc-Si core and the enhanced carrier lifetime by surface passivation. We anticipate that this "diffuse and transport" method could be applied in other scenarios for enhanced electron-hole extraction, e.g. from materials such as quantum dots (QD) that have low carrier mobility and lifetime.

In conclusion, we have reduced the surface recombination of SiNWs nearly two orders of magnitude by *in-situ* surface passivation using a thin layer of a-Si. The surface passivation of SiNWs increases the carrier diffusion length and lifetime by one and two orders of magnitude, respectively. The a-Si layer enables the core-shell nanowires to strongly absorb light of a broader range of wavelengths. The core-shell nanowire is 90-fold more sensitive when used a photodetector, compared to regular nanowires without surface passivation. We expect that the surface-passivated nanowires will have significantly larger energy conversion efficiency when used as solar cell devices.

References

1. Zhu, J.; Yu, Z. F.; Burkhard, G. F.; Hsu, C. M.; Connor, S. T.; Xu, Y. Q.; Wang, Q.; McGehee, M.; Fan, S. H.; Cui, Y. *Nano Letters* **2009,** 9, (1), 279-282.
2. Garnett, E.; Yang, P. D. *Nano Letters* **2010,** 10, (3), 1082-1087.
3. Hu, L.; Chen, G. *Nano Letters* **2007,** 7, (11), 3249-3252.
4. Seo K.; Wober M.; Steinvurzel P.; Schonbrun E.; Dan Y. P.; Ellenbogen T.; Crozier K. B. *Nano Letters* **2011,** 11, (4), 1851-1856.
5. Law, M.; Greene, L. E.; Johnson, J. C.; Saykally, R.; Yang, P. D. *Nature Materials* **2005,** 4, (6), 455-459.
6. Baxter, J. B.; Aydil, E. S. *Applied Physics Letters* **2005,** 86, (5).
7. Tsakalakos, L.; Balch, J.; Fronheiser, J.; Korevaar, B. A.; Sulima, O.; Rand, J. *Applied Physics Letters* **2007,** 91, (23).
8. Kempa, T. J.; Tian, B. Z.; Kim, D. R.; Hu, J. S.; Zheng, X. L.; Lieber, C. M. *Nano Letters* **2008,** 8, (10), 3456-3460.
9. Allen, J. E.; Hemesath, E. R.; Perea, D. E.; Lensch-Falk, J. L.; Li, Z. Y.; Yin, F.; Gass, M. H.; Wang, P.; Bleloch, A. L.; Palmer, R. E.; Lauhon, L. J. *Nature Nanotechnology* **2008,** 3, (3), 168-173.
10. Kelzenberg, M. D.; Turner-Evans, D. B.; Kayes, B. M.; Filler, M. A.; Putnam, M. C.; Lewis, N. S.; Atwater, H. A. *Nano Letters* **2008,** 8, (2), 710-714.
11. Stelzner, T.; Pietsch, M.; Andra, G.; Falk, F.; Ose, E.; Christiansen, S. *Nanotechnology* **2008,** 19, (29).
12. Demichel, O.; Calvo, V.; Besson, A.; Noe, P.; Salem, B.; Pauc, N.; Oehler, F.; Gentile, P.; Magnea, N. *Nano Letters* **2010,** 10, (7), 2323-2329.
13. Hasegawa, H.; Akazawa, M. *Applied Surface Science* **2008,** 255, (3), 628-632.
14. Moreira, M. D.; Venezuela, P.; Schmidt, T. M. *Nanotechnology* **2008,** 19, (6).
15. Garnett, E. C.; Yang, P. D. *Journal of the American Chemical Society* **2008,** 130, (29), 9224-+.
16. Tian, B.; Kempa, T. J.; Lieber, C. M. *Chemical Society Reviews* **2009,** 38, (1), 16-24.
17. Kelzenberg, M. D.; Boettcher, S. W.; Petykiewicz, J. A.; Turner-Evans, D. B.; Putnam, M. C.; Warren, E. L.; Spurgeon, J. M.; Briggs, R. M.; Lewis, N. S.; Atwater, H. A. *Nature Materials* **2010,** 9, (3), 239-244.
18. Fan, Z. Y.; Razavi, H.; Do, J. W.; Moriwaki, A.; Ergen, O.; Chueh, Y. L.; Leu, P. W.; Ho, J. C.; Takahashi, T.; Reichertz, L. A.; Neale, S.; Yu, K.; Wu, M.; Ager, J. W.; Javey, A. *Nature Materials* **2009,** 8, (8), 648-653.
19. Aberle, A. G. *Progress in Photovoltaics* **2000,** 8, (5), 473-487.
20. Nicollia.Eh. *Journal of Vacuum Science & Technology* **1971,** 8, (5), S39-&.
21. Chung, S. W.; Yu, J. Y.; Heath, J. R. *Applied Physics Letters* **2000,** 76, (15), 2068-2070.
22. Wu, Y. Y.; Yang, P. D. *Journal of the American Chemical Society* **2001,** 123, (13), 3165-3166.
23. Law, M.; Goldberger, J.; Yang, P. D. *Annual Review of Materials Research* **2004,** 34, 83-122.
24. Novotny, L.; Pohl, D. W.; Regli, P. *Journal of the Optical Society of America a-Optics Image Science and Vision* **1994,** 11, (6), 1768-1779.
25. Hosaka, S.; Shintani, T.; Miyamoto, M.; Kikukawa, A.; Hirotsune, A.; Terao, M.; Yoshida, M.; Fujita, K.; Kammer, S. *Journal of Applied Physics* **1996,** 79, (10), 8082-8086.
26. Gu, Y.; Kwak, E. S.; Lensch, J. L.; Allen, J. E.; Odom, T. W.; Lauhon, L. J. *Applied Physics Letters* **2005,** 87, (4), 3.
27. Klaassen, D. B. M. *Solid-State Electronics* **1992,** 35, (7), 961-967.
28. Cao, L. Y.; White, J. S.; Park, J. S.; Schuller, J. A.; Clemens, B. M.; Brongersma, M. L. *Nature Materials* **2009,** 8, (8), 643-647.

29. Cao, L. Y.; Fan, P. Y.; Vasudev, A. P.; White, J. S.; Yu, Z. F.; Cai, W. S.; Schuller, J. A.; Fan, S. H.; Brongersma, M. L. *Nano Letters* **2010,** 10, (2), 439-445.
30. Tsakalakos, L.; Balch, J.; Fronheiser, J.; Shih, M. Y.; LeBoeuf, S. F.; Pietrzykowski, M.; Codella, P. J.; Korevaar, B. A.; Sulima, O.; Rand, J.; Davuluru, A.; Rapol, U. *Journal of Nanophotonics* **2007,** 1.
31. Yoshida, N.; Shimizu, Y.; Honda, T.; Yokoi, T.; Nonomura, S. *Journal of Non-Crystalline Solids* **2008,** 354, (19-25), 2164-2166.
32. Serin, T.; Serin, N. *Applied Physics a-Materials Science & Processing* **1994,** 59, (4), 431-433.
**Acknowledgements**

This work is supported by Zena Technologies. Experiments were conducted in the Center for Nanoscale System (CNS) at Harvard University, which is supported by the National Science Foundation. We thank Dr. Jiangdong Deng for his assistance in setting up the scanning photocurrent microscopy system. We also thank Mr. Peter Duane for helping set up the monochromator system, Dr. Paul Steinvurzel for useful discussions on the calculation of the nanowire leaky mode resonances, and Mr. Kai Wang for discussion on the absorption cross-section of nanowires.

**Additional information**

The authors declare no competing financial interests. Correspondence and requests for materials should be addressed to Kenneth B. Crozier (kcrozier@seas.harvard.edu).

**Figure Captions**

**Figure 1 Transmission electron microscope (TEM) images of a core-shell silicon nanowire.** (a) High-resolution TEM image, showing that core is single-crystalline while shell is amorphous. Scale bar, 3 nm. (b) TEM image, showing that amorphous shell is ~10nm thick and crystalline core is ~40nm in diameter. Inset: energy dispersive X-ray spectroscopy (EDS) indicates that amorphous shell is silicon. Scale bar, 30 nm.

**Figure 2 Current-voltage characteristics and scanning photocurrent microscopy (SPCM) of core-shell silicon nanowires** The nanowires are intentionally contacted with two Schottky barriers. (a) I-V measurement (dark current, black curve) shows that, under positive or negative voltage, one barrier is always under reverse bias. I-V measurement with illumination (red curve) is obtained with white light from microscope lamp. Inset: transimpedance measurement, indicating that nanowire is p-type. (b) Conceptual illustration of method for measuring carrier diffusion lengths. (c) Nanowire topography, obtained by AFM functionality of near-field scanning optical microscope. Scale bar, 1μm. (d) Photocurrent map, recorded simultaneously with topographic image. Scale bar, 1μm. (e) Photocurrent current profiles along nanowire axis (blue dashed line of panel d) for different voltages applied to nanowire. Plotted in logarithmic scale, slope of photocurrent vs distance curve does not change with bias voltage, indicating that the excited charge carriers are diffusive. Larger value of illumination spot location corresponds to it being closer to Schottky contact. (f) At conditions similar

to (e), photocurrent profiles along axes of nanowires with diameters of 190nm, 77nm and 51nm.

**Figure 3 Surface passivation effects in nanowires of different diameters.** Diffusion lengths vs nanowire diameter, for core-shell (black squares) and regular nanowires without shell (red dots and green stars). Shell thickness is excluded from the value of the diameter because carriers mainly transport in the core, as data show later. Red dots represent data from ref. 9. Green stars are from measurements we obtain on nanowires without shell. Black lines are calculated from *equation (1)*, assuming surface recombination velocities S of $3\times10^5$, $1\times10^4$, $4.5\times10^3$, and $3\times10^3$ cm/s, with diffusion coefficient D of 4.8 cm$^2$/s and bulk recombination lifetime $\tau_b$ extracted to be ~2 ns.

**Figure 4 Diagram of experimental set-up and measured AC photosensitivity spectra.** (a) Four point probe measurement technique is used, with $V_s$ being the power supply voltage and $V_{in}$ the measured voltage. Bottom right: nanowire cross section is shown together with definitions of TE and TM illumination. (b) photoconductance of a core-shell nanowire at different illumination intensities. Inset: photocurrent linearly increases with the bias $V_{in}$. (c) Measured photosensitivity vs wavelength for a core-shell nanowire (black squares) and a nanowire without shell (red dots). The photosensitivity for the nanowire without shell has been magnified 50-fold for comparison purposes. The core-shell SiNW has a core diameter of ~70 nm, and an outer diameter of ~90 nm. The nanowire without shell is 130 nm in diameter. FDTD-simulated absorption cross-section is plotted for the core-shell nanowire (black line) and the nanowire without shell (red line). (d) Simulated electric field intensity ($E^2$, normalized to incident illumination) of leaky modes for TM illumination. The nanowires are on quartz

substrates. The outer circle is the wire-air interface while the inner circle is the interface between a-Si and sc-Si. Upper plot is for the core-shell nanowire. Middle and lower plots are for the regular nanowire without shell.

**Figure 5 Absorption cross-section and carrier transport model of the core-shell nanowire.** (a) Simulated absorption cross-section vs wavelength for the core-shell nanowire in Fig. 4c. Total absorption cross-section, i.e. of core and shell, is plotted as red curve. Absorption cross-section of core is plotted as black curve. (b) Charge carrier transport model. Excess carriers excited by the strong absorption in the shell diffuse into the core where they are efficiently transported to the electrodes as photocurrent.

**Figures**

**Figure 1**

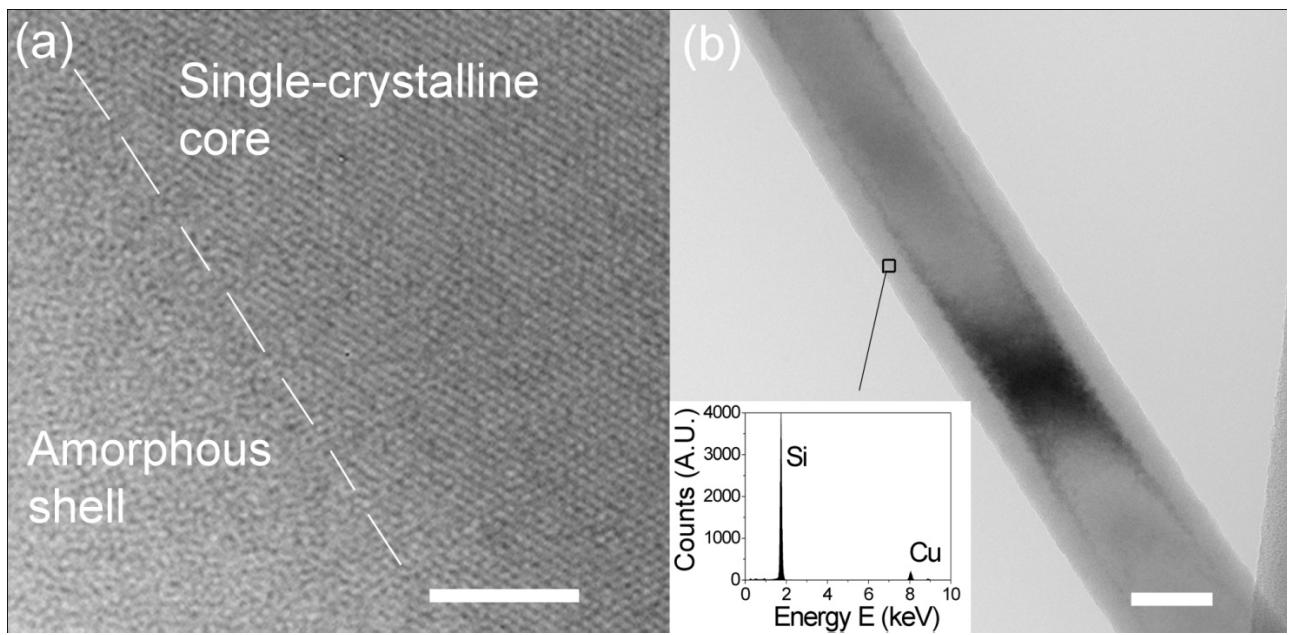

**Figure 2**

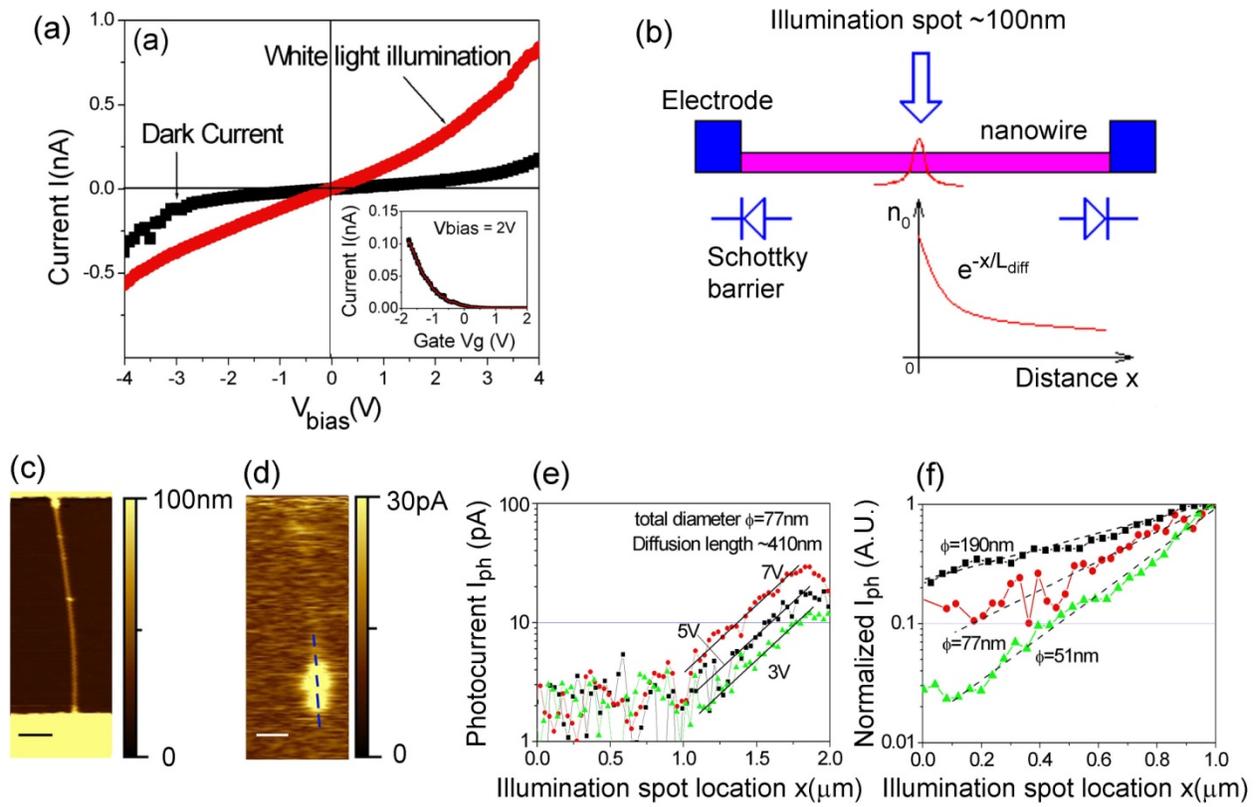

**Figure 3**

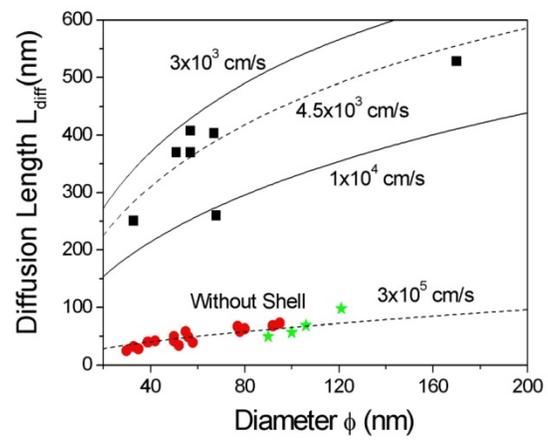

**Figure 4**

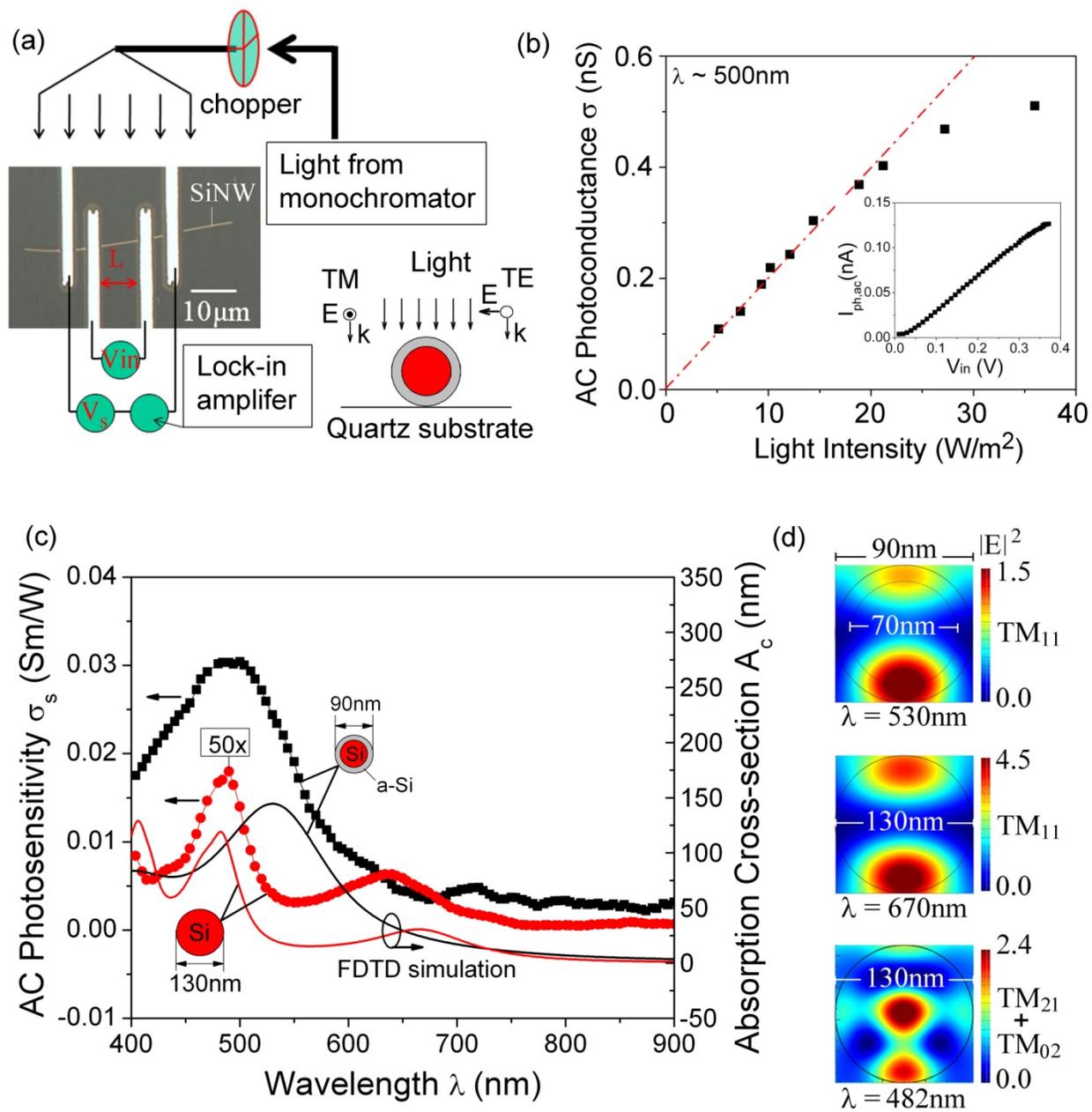

**Figure 5**

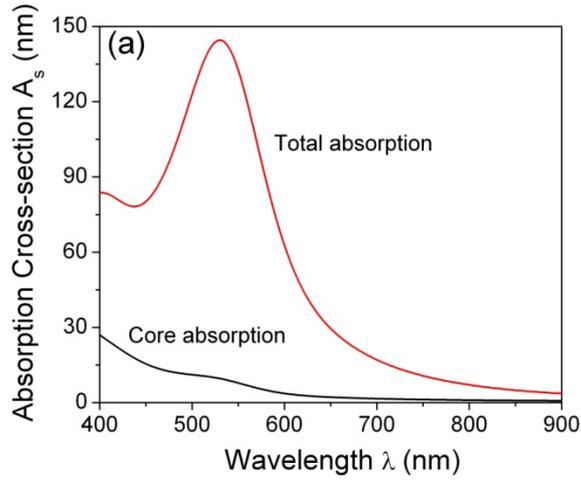
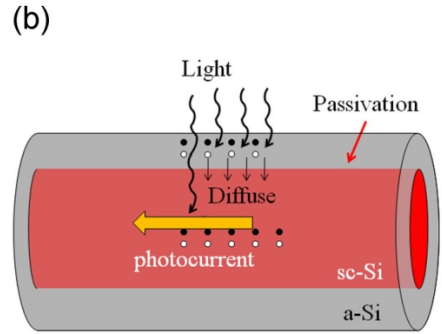

**TOC**

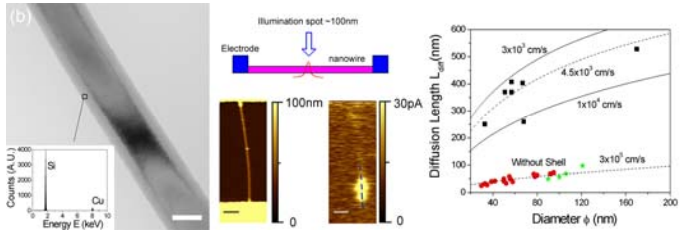